\documentstyle[preprint,aps]{revtex}
\begin{document}
\preprint{}
\draft
\title{Two body non-leptonic $\Lambda_b$ decays
in the large $N_c$ limit}
\author{A. K. Giri, L. Maharana and R. Mohanta}
\address{Department of Physics, Utkal
University, Bhubaneswar-751004, India.} 
\maketitle
\begin{abstract}
The two body non-leptonic $\Lambda_b$ decays
are analyzed in the HQET with factorization approximation
and large $N_c$ limit. In this limit, $\Lambda_b$ and
$\Lambda_c$ baryons can be treated as the bound states of 
chiral soliton and heavy meson, and consequently the Isgur-Wise function 
comes out in a straight forward manner. The results obtained 
remain well below their previously predicted  upper limit.\\
\\
Key Words : Heavy Quark Effective Theory, Factorization Approximation
\end{abstract}

\pacs{PACS : 11.30.Hv, 13.30.Eg}

\section{Introduction}

In the last few years cosiderable progress has been achieved
in the understanding of the weak decays of heavy hadrons due
to the development of the heavy quark effective theory (HQET)
[1-5].
Contrary to the significant progress made in the studies of 
the meson decays, advancement in the arena of heavy baryons 
has been very slow. In particular
the non-leptonic weak decays
of heavy baryons have not been understood clearly till now. 
At present, there are not many experimental results available
for heavy baryons. But in the future we may expect more and 
more data coming from the colliders. Hence the study of heavy
baryons is of great interest in the near future. 
In fact some phenomenological approaches such as pole model, current
algebra etc. have been employed to analyze these decay
processes. The well known factorization hypothesis
[6-12] which has been applied successfully to the 
heavy meson decays can also be applied for heavy baryon cases
for large $N_c$ limit [13].

In this article we intend to study the 
two body non-leptonic $\Lambda_b$
decays in the heavy quark effective theory [1-5] 
considering factorization approximation.
The HQET provides a convenient and simplified framework to analyze the
weak decays of heavy hadrons, composed of one heavy quark
and any number of light quarks. Of particular importance
are the semileptonic decays of heavy mesons, where in the limit
of infinite quark masses, all the hadronic form factors
can be expressed in terms of a single universal function
$\xi(v\cdot v^\prime)$, the Isgur-Wise function [1]. The 
function depends only on the four velocities of the heavy 
particles involved, and is normalized at the point of zero recoil.
Similarly in case of weak decays of heavy
baryons one can also write the form factors in terms of another 
Isgur-Wise function [3, 5].

We therefore consider it worthwhile to investigate the 
two-body nonleptonic decays $\Lambda_b \rightarrow \Lambda_c^+
P^-$ and $\Lambda_b \rightarrow \Lambda_c^+ V^-$, 
where $P$ and $V$ denote pseudoscalar and vector mesons respectively,
using the HQET in conjunction with factorization 
approximation. In fact factorization method works well
for the description of non-leptonic decays of heavy baryons,
in the large $N_c$ limit [13]. The use of the
HQET implies that the expression for the decay widths should
contain the universal Isgur-Wise function, which is normalized
at the point of zero recoil. However other than at the point of 
zero recoil, HQET does not predict the shape of the Isgur-Wise
(IW) function. Therefore to know the value of Isgur-Wise function 
at the particular
kinematic point of interest, we evaluate it in the
large $N_c$ limit [14, 15], where factorization approximation
is valid. 
Earlier these decays have been studied by Mannel 
et al. [13]. They have parametrized the Isgur-Wise form factor
$G_1(v\cdot v^\prime)$ in three different forms as :
\begin{eqnarray}
&&G_1(v\cdot v^\prime)=1+\frac{1}{4}a (v-v^\prime)^2 (v+v^\prime)^2
\;,\;\;\; G_1(v\cdot v^\prime)=\frac{1}{1-(v-v^\prime)^2/\omega_0^2}\;,
\nonumber\\
&&\mbox{and}\;\;\;\;\;G_1(v\cdot v^\prime)=\mbox{exp} [b(v-v^\prime)^2]\;,
\end{eqnarray}
and they have taken the values of the parameters
$a$, $b$, and $\omega_0$,
from the work on $B$ meson decays [16]. 
They have  also predicted the upper limits for the branching
ratios for these decay processes, considering the normalized value
of the IW function.

The report is organized as follows. In Section II we present the
general framework for the study of the nonleptonic decays in
the factorization method. The Isgur-Wise function is evaluated 
in the large $N_c$ limit in section III. Section IV 
contains results and discussions.

\section{General Framework}

Neglecting the penguin contribution, the four fermion effective
Hamiltonian relevant to the $\Lambda_b \rightarrow \Lambda_c^+
P^-$ and $\Lambda_b \rightarrow \Lambda_c^+
V^-$ decays is given as [13] 
\begin{equation}
{\cal H}_{eff}=\frac{G_F}{\sqrt 2}V_{UD}^*V_{cb}\;[C_1(m_b)
{\bf {O_1}}+C_2(m_b){\bf {O_2}}] , 
\end{equation}
with
\begin{equation}
{\bf {O_1}}=(\bar D U)^\mu(\bar c b)_\mu \;\;
\mbox{and} \;\;{\bf {O_2}}=
(\bar c U)^\mu(\bar D b)_\mu,\;
\end{equation}
where $G_F$ is the fermi coupling constant and the quark current
$(\bar q^\prime q)_\mu$ is a short hand for $\bar {q}^\prime_\alpha
\gamma_\mu(1-\gamma_5)q_\alpha$;
$\alpha$ is the color index. 
$U$ and $D$ are either $c$, $s$ or $u$, $d$ quarks. Thus for $U$, $D$
= $u$, $d$ we have
$\pi^-$ and $\rho^-$ in the final state as $P$ and $V$ while
for $U$, $D$ = $c$, $s$ the final $P$/$V$ states are $D_s$/$D_s^*$ mesons. 
The values of the Wilson
coefficients $C_{1,2}$ can be calculated using the Leading
Logarithmic Approximation (LLA) [17] and are given as

\begin{equation}
C_1(m_b)=1.11 \;\;\; \mbox{and} \;\;\;C_2(m_b)=-0.26 .
\end{equation}

The decays  $\Lambda_b \rightarrow \Lambda_c^+ P^-$ and
$\Lambda_b \rightarrow \Lambda_c^+ V^-$ 
can occur
by the operator ${\bf O_1}$ where it is assumed that the $\Lambda_b
\rightarrow \Lambda_c $ transition is caused by the current
operator $(\bar c b)$ and that $P^-(V^-)$ are created by the current
operator $(\bar D U)$. In the factorization approximation it is 
assumed that the $\Lambda_b\rightarrow
\Lambda_c$ transition and the $P^-(V^-)$ creation are
independent of each other,
and hence the amplitude can be written as
\begin{equation}
<\Lambda_c^+ P^-(V^-)|(\bar D U)(\bar c b)|\Lambda_b>=
<P^-(V^-)|(\bar D U)|0><\Lambda_c^+|(\bar c b)|\Lambda_b>.\;
\end{equation}
In the large $N_c$ limit, where factorization approximation is
valid, the contribution of ${\bf O_2}$ to these decays
is suppressed.
Therefore one can write 
the amplitude for the decays $\Lambda_b \rightarrow
\Lambda_c^+ P^-$ and $\Lambda_b \rightarrow \Lambda_c^+ V^-$ as
\begin{equation}
{\cal M} (\Lambda_b \rightarrow \Lambda_c^+P^-(V^-))=\frac{G_F}{\sqrt
2} V_{UD}^* V_{cb} C_1(m_b)<P^-(V^-)|(\bar DU)^\mu|0>
<\Lambda_c^+|(\bar cb)_\mu|\Lambda_b>.\;\label{eq:eq1}
\end{equation}
To evaluate the factorized amplitudes we use the following
matrix elements. 
\begin{equation}
<P(p)|(\bar DU)^\mu|0>=-i f_P\; p^\mu ,
\end{equation}
and
\begin{equation}
<V(p,\epsilon)|(\bar DU)^\mu|0>= f_V\; M_V \;\epsilon^\mu ,
\end{equation}
where $f_P$ and $f_V$ are the pseudoscalar and vector meson
decay constants respectively. The matrix element 
$<\Lambda_c|(\bar cb)_\mu|\Lambda_b>$ is given in the HQET [3, 4] 
as
\begin{equation}
<\Lambda_c^+ (v^\prime, s^\prime)|\bar c \gamma_\mu (1-\gamma_5)b|
\Lambda_b (v,s)>=\eta (v\cdot v^\prime)\;\bar u_c(v^\prime, s^\prime) 
\gamma_\mu
(1-\gamma_5)u_b(v,s)\;,\label{eq:eq2}  
\end{equation}
where $\eta(v\cdot v^\prime)$ is the baryonic Isgur-Wise
function, $u_c(v^\prime, s^\prime)$
and $u_b(v, s)$ are the spinors of the $\Lambda_c$ and $\Lambda_b$
baryons. Thus with Eqs. (\ref{eq:eq1}-\ref{eq:eq2}) 
we obtain the decay widths for the decay processes $\Lambda_b \rightarrow 
\Lambda_c^+ P^-$ and $\Lambda_b \rightarrow 
\Lambda_c^+ V^-$, given as

\begin{eqnarray}
\Gamma(\Lambda_b(v) & \rightarrow & \Lambda_c^+(v^\prime) P^-(p))=
\frac{G_F^2}{8\pi M_{\Lambda_b}^2}|V_{UD}^*V_{cb}|^2 \;C_1^2(m_b)\; f_P^2
\;\eta^2(v\cdot v^\prime)\;|\vec p|\nonumber\\
&&\times [(M_{\Lambda_b}^2-M_{\Lambda_c}^2)^2-M_P^2(M_{\Lambda_b}^2+
M_{\Lambda_c}^2)],\;\label{eq:eq11}
\end{eqnarray}
and
\begin{eqnarray}
\Gamma(\Lambda_b(v) &\rightarrow &\Lambda_c^+(v^\prime) V^-(p))=
\frac{G_F^2}{8\pi M_{\Lambda_b}^2}|V_{UD}^*V_{cb}|^2 \;C_1^2(m_b)\; f_V^2
\;\eta^2(v\cdot v^\prime)\;|\vec p|\nonumber\\
&&\times [(M_{\Lambda_b}^2-M_{\Lambda_c}^2)^2+M_V^2(M_{\Lambda_b}^2+
M_{\Lambda_c}^2-2M_V^2)],\;\label{eq:eq12}
\end{eqnarray}
where $|\vec p|$ is the c.o.m mementum of the emitted particles in
the rest frame of initial $\Lambda_b$ baryon and $M$'s are the 
corresponding pseudoscalar, vector meson and 
baryon masses. The above expressions for the decay widths 
contain besides the known
quantities, the unknown Isgur-Wise function, which can be
calculated in the large $N_c$ limit in a simple manner.
 
\section{Evaluation of the Isgur-Wise Function}

Here we have presented the evaluation of the Isgur-Wise function
in the same manner as suggested in Ref. [14]. In the large $N_c$
limit the light baryons n, p, $\Delta$ etc. can be viewed as
solitons in the chiral Lagrangian for pion self interaction
[18]. The baryons containing a single heavy charm (or bottom)
quark are bound states of these solitons with $D$ and $D^*$
(or $B$ and $B^*$) mesons [19-22]. In this paper we use the
bound state soliton picture to estimate the value of the 
baryonic Isgur-Wise function. In the ground state of $\Lambda_Q$
baryons, the light quarks are in the spin $0$ state
[23]. Hence in the bound state soliton picture, $\Lambda_Q$-type
bound state arise when the spin of the light degrees of freedom
of the heavy meson and the spin of the nucleon are combined into
a spin zero configuration where as the isospin of the heavy meson
and that of the nucleon are combined into an isospin zero state.
Other baryons (e.g, the $\Delta$) only contribute to the bound states
with higher isospin.

Let the light degrees of freedom of the heavy baryon is denoted
by $|I,I_3;s_l,m_l>$, where $I$ and $s_l$ denote their isospin
and spin quantum numbers while $I_3$ and $m_l$ are their third 
components respectively. Hence the light degrees of freedom
of $\Lambda_Q$ baryon is denoted by $|0,0;0,0>$. The chiral
soliton is denoted by $|R,b;R,n)$, where $R=1/2$ for the
nucleon. On the other hand the light degrees of freedom of the 
heavy meson is given as $|1/2,c;1/2,p\}$.

In the large $N_c$ limit, the binding potential between the
chiral soliton and heavy meson is independent of both the
isospin and spin of the particles. Hence for the light degrees
of freedom of $\Lambda_Q$ baryon, we have the decomposition as
\begin{eqnarray}
|0,0;0,0 \;(v)>&&=\int d^3{\bf q}\; \Phi_Q({\bf q})\; (1/2,b;1/2,c|0,0)
\;(1/2,n;1/2,p|0,0)\nonumber\\
\times &&|1/2,b;1/2,n\;(-{\bf q}+M_B{\bf v}))\;|1/2,c;1/2,p\;
({\bf q}+M_H{\bf v})\},
\end{eqnarray} 
where $(j_1,m_1;j_2,m_2|J,M)$'s are the Clebsch-Gordan
coefficients. $\Phi_Q({\bf q})$ is the ground state wave function,
$M_B$ and $M_H$ are the masses of the chiral soliton and
heavy meson respectively. 

The spin-1/2 $\Lambda_Q$ baryon is composed of a spin-1/2 heavy
quark and spin-0 light degrees of freedom.
Hence the matrix element of the current $\bar c\gamma_\mu
(1-\gamma_5)b$ between $\Lambda_b$ and $\Lambda_c$ baryons
is given as
\begin{equation}
<\Lambda_c^+(v^\prime, s^\prime)|\bar c\gamma_\mu(1-\gamma_5)b|
\Lambda_b(v, s)>=<0,0;0,0\;(v^\prime)|0,0;0,0\;(v)>
\bar u_c\gamma_\mu(1-\gamma_5)u_b .\label{eq:eq3}
\end{equation}
Comparing  Eqns. (\ref{eq:eq2}) and (\ref{eq:eq3}) 
we obtain the expression for 
the baryonic IW function as
\begin{eqnarray}
\eta(v\cdot v^\prime)&=&<0,0;0,0\;(v^\prime)|0,0;0,0\;(v)>\nonumber\\
&=&\int d^3 {\bf {q^\prime}}\;d^3 {\bf q}\; \Phi_c^*({\bf {q^\prime}})\;
\Phi_b({\bf q})\nonumber\\ 
&&\times(1/2,b^\prime;1/2,c^\prime|0,0)^*\;
(1/2,n^\prime;1/2,p^\prime|0,0)^*\; (1/2,b;1/2,c|0,0)\;
(1/2,n;1/2,p|0,0)\nonumber\\ 
&&\times(1/2,b^\prime;1/2,n^\prime\;(-{\bf
q^\prime}+M_B{\bf v}^\prime)\; |1/2,b;1/2,n\;(-{\bf q}
+M_B{\bf v}))\nonumber\\
&&\times\{1/2,c^\prime;1/2,p^\prime\;({\bf q^\prime}+M_H{\bf
v}^\prime)\;| 1/2,c;1/2,p\;({\bf q}+M_H{\bf v})\}.\label{eq:eq4}
\end{eqnarray}
Using the normalization conditions for the chiral soliton and 
heavy meson states, it is found that all the 
Clebsch-Gordan coefficients in (\ref{eq:eq4}) turn out to be 
unity and the Isgur-Wise function is given as
\begin{equation}
\eta(v\cdot v^\prime)=\int d^3{\bf q}\;\Phi_c^*({\bf q})\;
\Phi_b({\bf q}+ M_B({\bf v}-{\bf v}^\prime)).\label{eq:eq7}
\end{equation}
It is noted from Eqn. (\ref{eq:eq7}) that the IW function depends on the 
spatial wave function $\Phi_Q(\bf{q})$ of the $\Lambda_Q$
baryon. In the large $N_c$ limit, the binding potential between
the heavy meson and the chiral soliton is simple harmonic [15],
and hence the wave function is taken as
\begin{equation}
\Phi_Q({\bf q})=\frac{1}{(\pi^2\mu_Q\kappa)^{3/8}}\;
\mbox{exp} \left(-\frac{{\bf q}^2}{2\sqrt{\mu_Q\kappa}}\right),\label{eq:eq8}
\end{equation}
where  $\mu_Q=M_B M_H/(M_B +M_H)$, is the reduced mass of the
bound state, $M_H$ denotes the masses of $D/B$ mesons for 
$\Phi_c$/$\Phi_b$ wave functions.
$\kappa$ is the spring constant and its value is taken to be
$(440 \;MeV)^3$ [24]. In the rest frame of the initial
state, $v=(1,\vec 0)$ and ${\bf v}^\prime $ directed
along z-axis we obtain the Isgur-Wise function (\ref{eq:eq7}) using 
(\ref{eq:eq8}) for non-relativistic recoils i.e., $|{\bf v}^\prime|^2
\approx 2\;(v\cdot v^\prime -1)$, as 

\begin{equation}
\eta(v\cdot v^\prime)= \left [ \frac{4 \sqrt{\mu_b
\mu_c}}{(\sqrt{\mu_b}+ \sqrt{\mu_c})^2}\right ]^{3/4}\;
\mbox{exp} \left (-\frac{(v\cdot v^\prime -1) M_B^2}{\sqrt{\kappa}
(\sqrt{\mu_b} +\sqrt{\mu_c})}\right ).\label{eq:eq9}
\end{equation}
It should be noted from eqn. (17) that the Isgur-Wise
function slightly deviates from unity at the point of zero recoil.
This violation of normalization condition can be explained 
as follows.
The heavy quark symmetry arises 
in the limit of QCD, where
the heavy quark mass $m_Q$ is taken formally to be infinite
and in this limit all the hadronic form factors can be expressed 
in terms of the IW function. However here we have used finite
masses for the heavy mesons (i.e., $B$ and $D$ mesons), and hence heavy
quark symmetry breaks down. Thus breaking of the heavy flavor symmetry
causes a violation of the normalization condition $\eta(1)=1$.

The product $(v\cdot v^\prime)$ is determined by considering the
kinematics of the system. Since we are dealing with the two body
decays $\Lambda_b(v) \rightarrow
\Lambda_c^+(v^\prime)P^-(p)/V^-(p)$, from momentum conservation we
obtain  
\begin{equation}
v\cdot v^\prime =\frac{M_{\Lambda_b}^2+M_{\Lambda_c}^2-M_{P/V}^2}
{2 M_{\Lambda_b}M_{\Lambda_c}} .\label{eq:eq10}
\end{equation}
Taking the masses of the particles from Ref. [25]
the values of the Isgur-Wise functions are
calculated with eqns. (\ref{eq:eq9}) and (\ref{eq:eq10}) 
as presented in Table-1.

\section{Results and Discussions}

Having obtained the values of the Isgur Wise function,
we use the following data to estimate the decay widths for the 
processes $\Lambda_b \rightarrow \Lambda_c^+ P^-$ and
$\Lambda_b \rightarrow \Lambda_c^+ V^-$ using eqns. (10)
and (11).
The CKM matrix elements $V_{cb}$=0.041,
$V_{ud}$=0.0976 and $V_{cs}$=0.9743 are taken from Ref. [25] and
the values of the decay constants used are $f_{\pi}$=130.7 MeV,
$f_{D_s}$=$f_{D_s^*}$ = 232 MeV [25] and $f_\rho$ =210 MeV [13].
With these values we have evaluated the
branching ratios for several two body non-leptonic
$\Lambda_b$ decays as presented in Table-1.

In this work we have estimated the branching ratios for
$\Lambda_b \rightarrow \Lambda_c^+ P^-$ and
$\Lambda_b \rightarrow \Lambda_c^+ V^-$ decays in the heavy quark
effective theory with factorization 
approximation. In fact
factorization method works well for the description of
non-leptonic decays of heavy baryons in the large $N_c$ limit
[13]. The use of HQET allows us to write the weak decay form
factors in terms of the Isgur-Wise function. However it does not 
predict the shape of the IW function except at the point of
zero recoil, where it is normalized to unity. Therefore to know
the value of Isgur-Wise function at the particular kinematic
point of interest, we have evaluated it in the large $N_c$
limit [14, 15]
considering
the bound state soliton picture.
These decays have been previously studied in Ref. [13]
in which they
have parametrized the Isgur-Wise function 
in three different forms (eqn.(1)) and used the values of the unknown
parameters from the work on $B$ meson decays [16]. However in our case
we have evaluated the Isgur-Wise function in the large $N_c$ limit
where factorization approximation is valid
and the results came out in a straight forward manner. 
Therefore our predicted 
results differ from theirs as noted from Table-1.
They have also predicted 
the upper limit of the branching fractions for these decays, by
considering the normalized value of the Isgur-Wise function. 
The results of
the present investigation lie well below their corresponding upper
limit. As the
experimental data are expected in the future from the 
Colliders, so these results can be verified, which will
definitely enrich our understanding in this sector to a
greater extent.

\begin{table}
\caption{Prediction for the branching ratios $BR$ (in \%) for the two body 
nonleptonic $\Lambda_b$ decays in the large $N_c$ limit.}
\begin{tabular}{cccccc}
\tableline
&&& $BR$ & $BR$ & ${BR}_{max}$.\\
Decay Process & $|\vec p|$ in MeV & $\eta(v\cdot v^\prime)$ &
Present calculation & Ref. [13]& Ref. [13] \\
\tableline
$\Lambda_b \rightarrow \Lambda_c^+\pi^-$ & 2355.343 &
0.456 & 0.342 & $0.46_{-0.31}^{+0.20}$ & 2.0 \\
$\Lambda_b \rightarrow \Lambda_c^+D_s^-$ &1849.748 &
0.596 & 1.156 & $2.3_{-0.40}^{+0.30}$& 6.5 \\
$\Lambda_b \rightarrow \Lambda_c^+\rho^-$ & 2284.288 &
0.474 & 0.954 & $0.66_{-0.40}^{+0.24}$& 2.5 \\
$\Lambda_b \rightarrow \Lambda_c^+D_s^{*-}$ & 1765.854 &
0.621 & 1.769 &$1.73_{-0.30}^{+0.20}$& 4.7
\end{tabular}
\end{table}

\end{document}